\documentclass{article}
\usepackage{graphicx}
\usepackage{geometry}
\begin{document}
\title{The QCD equation of state at nonzero
densities: lattice result} 
\author{
Z.~Fodor$^{a}$, S.D.~Katz$^b$ and K.K. Szab\'o$^{a}$\\
 \it $^a$Institute for Theoretical Physics, E\"otv\"os University, P\'azm\'any
1, H-1117 Budapest, Hungary\\
\it $^b$Deutsches Elektronen-Synchrotron DESY, Notkestr. 85, D-22607,
Hamburg, Germany}

\date{\today}
\maketitle
\begin{abstract}
In this letter we give the equation of state of QCD at 
finite temperatures and
densities. The recently proposed overlap improving multi-parameter
reweighting technique is used to determine observables at nonvanishing
chemical potentials. Our results
are obtained by studying $n_f$=2+1 dynamical
staggered quarks with semi-realistic masses on $N_t=4$ lattices.
\end{abstract}

\vspace*{-8.0cm}
\noindent
\hfill 
\vspace*{7.8cm}

\vspace*{0.2cm}

According to 
the standard picture of QCD, at high temperatures 
and/or high densities there is
a change from a state dominated by hadrons
to a state dominated by partons. This transition
happened in the early Universe (at essentially 
vanishing density) and probably happens in heavy ion 
collisions (at moderate but non-vanishing density) 
and in neutron stars (at large density, for which
a rich phase structure is conjectured
\cite{Alford:1997zt,Alford:1998mk,Rapp:1997zu,Rajagopal:2000wf}).
There are well established nonperturbative lattice 
techniques to study this 
transition at vanishing density, at which
the equation of state were
determined as a function of the temperature.
Due to the sign problem (oscillating signs lead to cancellation
in results, a phenomena which appears in many fields of 
physics) nothing could have been said about the experimentally/observationally relevant case at non-vanishing
densities.
By using our recently proposed lattice technique \cite{Fodor:2001au} 
we calculate the equation of state 
at non-vanishing temperatures and densities.
In this letter we
present the technique and show the results. A detailed version
of the present work will be published elsewhere \cite{Csikor:2002aa}.

\begin{figure}
\begin{center}
\includegraphics*[width=8.5cm,bb=0 100 413 373]{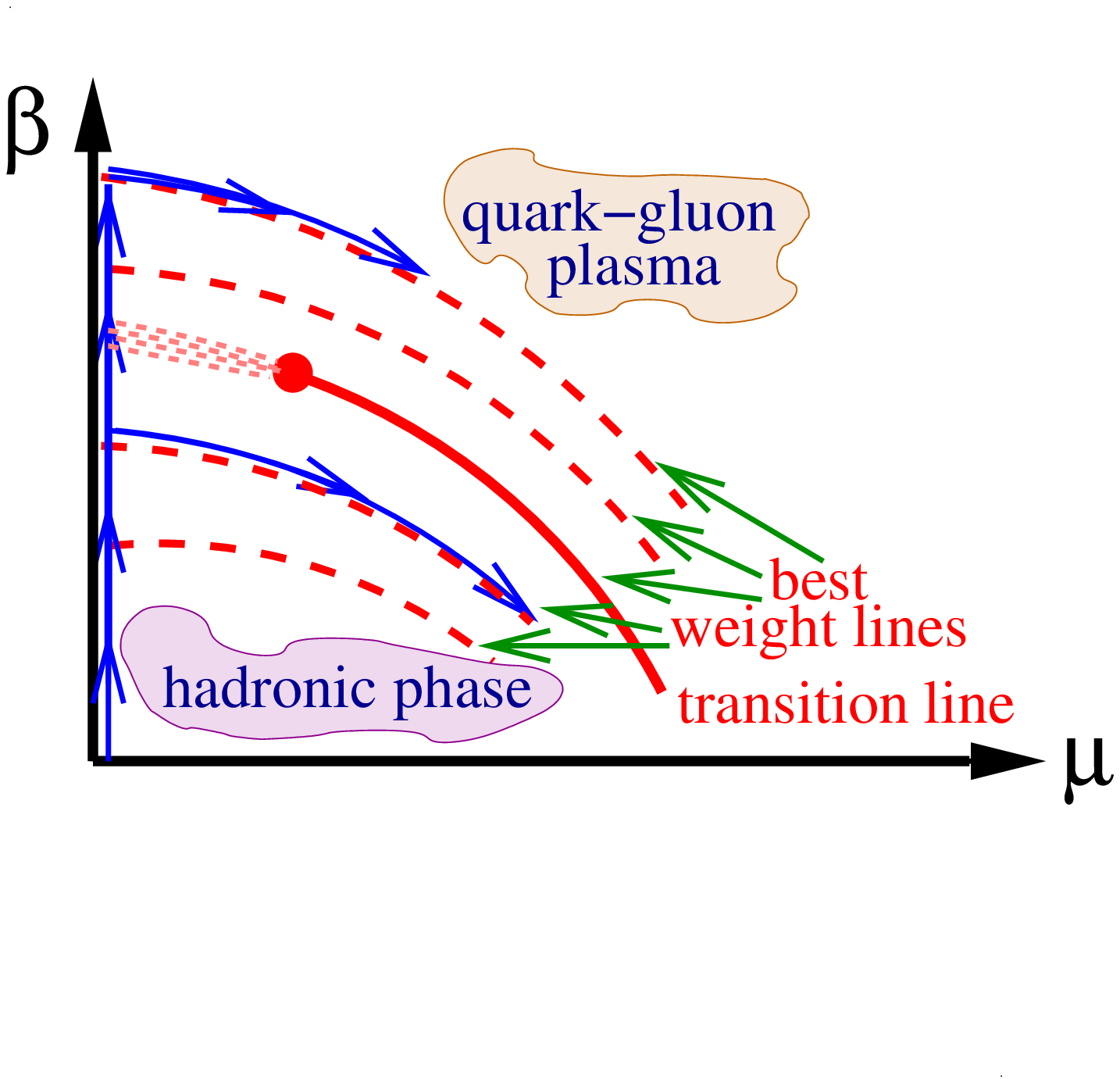}
\end{center}
\caption{\label{weightlines}
The best weight lines on the $\mu$--$\beta$ plane, 
along which reweighting is performed. In the middle we 
indicate the transition line. Its first dotted part
is the crossover region. The blob represents the 
critical endpoint, after which the transition is of first order. 
Below the transition line the system is in the hadronic phase,
above the transition line we find the QGP. 
The integration paths used to calculate the pressure are shown by the 
arrows along the $\mu=0$ axis and the best weight lines.
}
\end{figure}

QCD at nonzero density 
can be studied by introducing the
chemical potential ($\mu$). Such a theory is 
easily formulated on space-time lattice
\cite{Hasenfratz:1983ba,Kogut:1983ia,Montvay:1994aa}.
However; standard, importance sampling based 
Monte-Carlo techniques can not be used 
at $\mu\neq$0.  
Up to now, no technique was suggested capable of giving
the equation of state (EOS)
at non-vanishing $\mu$, which would be essential 
to describe the quark gluon plasma (QGP) formation
at heavy ion collider experiments. Results are only available for $\mu$=0
\cite{Gottlieb:1996ae,Karsch:2000ps,AliKhan:2001ek} at
non-vanishing temperatures (T) (a recent review is Ref. \cite{Ejiri:2000bw}). 

The overlap improving
multi-parameter reweighting \cite{Fodor:2001au} opened the possibility
to study lattice QCD
at nonzero $T$ and $\mu$. First one produces
an ensemble of QCD configurations at $\mu$=0 and at T$\neq$0.
Then the Ferrenberg-Swendsen type reweighting factors
\cite{Ferrenberg:yz} of these configurations are determined
at $\mu\neq 0$ and at a lowered T. The idea can be
easily expressed in terms of the partition function
\begin{eqnarray}\label{reweight}
Z(\mu,\beta) = \int {\cal D}U\exp[-S_{g}(\beta,U)]\det M(\mu,m,U)= 
\nonumber \\
\int {{\cal D}U \exp[-S_{g}(\beta_0,U)]\det M(\mu=0,m,U)}
\nonumber \\
{ \left\{\exp[-S_{g}(\beta,U)+S_{g}(\beta_0,U)]
\frac{\det M(\mu,m,U)}{\det M(\mu=0,m,U)}\right\} },
\end{eqnarray}
where $S_g$ is the action of the gluonic field ($U$), 
$\beta=6/g^2$ fixes the coupling of the strong interactions ($g$). 
Note that for a given lattice $T$
is an increasing function of $\beta$. The quark mass parameter is $m$ and
$\det M$ comes from the integration over the quark fields. At nonzero 
$\mu$ one gets a complex $\det M$ which has no probability interpretation, thus
it spoils any importance sampling.
Therefore, the first expression of eq. (\ref{reweight}), $\mu\neq$0,
is rewritten in a way that the second line of eq. (\ref{reweight})
is used as an integration measure (at $\mu$=0, for which 
importance sampling works) and the remaining part in the curly
bracket is measured on each independent configuration and interpreted 
as a weight factor $\{ w(\beta,\mu,m,U)\}$. 
In order to maximise the accuracy of 
$Z$ the reweighting is performed along the best weight lines on the 
$\mu$--$\beta$ plane (or equivalently on the $\mu$--$T$ plane). 
These best weight lines are defined by minimising the spread of $\log w$. 

Similar reweighting can 
be done in the mass parameter, too. 
Using the above technique, transition (or hadronic/QGP) 
configurations are reweighted to transition
(or hadronic/QGP) configurations as illustrated by Fig. \ref{weightlines};
thus, a much better overlap can be obtained than by 
reweighting pure hadronic
configurations to transition ones as done 
by single $\mu$-reweighting
\cite{Barbour:1997ej}. 
As we emphasised, the technique works for
temperatures at, below and above the transition 
temperature ($T_c$). 
By using the reweighting technique, 
the phase diagram \cite{Fodor:2001au} and the location
of the critical endpoint \cite{Fodor:2001pe} was given
(for other approaches see e.g. \cite{Berges:1998rc,Halasz:1998qr}). 
Using a Taylor expansion around $\mu$=0, T$\neq$0 for small $\mu$ is
a variant of our multi-parameter reweighting method, which can be used
to determine thermal properties \cite{Allton:2002zi}. 
A completely
different method, analytic continuation from imaginary
$\mu$, confirmed the result of the
reweighting technique on the phase 
diagram \cite{deForcrand:2002ci}. 

In the present analysis we use $N_t\cdot N_s^3$ finite $T$ lattices
with $N_t$=4 and $N_s$=8,10,12 for reweighting 
and we extrapolate to the thermodynamical
limit using the available volumes ($V$). 
At $T$=0 lattices of $24\cdot14^3$ 
are taken for vacuum subtraction and
to connect lattice parameters to physical
quantities. 16 different $\beta$ values are used, which
correspond to $T/T_c=0.8,\dots,3$ and at $T_c$ to a lattice
spacing ($a$) of $\approx$0.28~fm. We use 2+1 flavours of
dynamical staggered quarks. 
Only $\mu$ values of the light quarks are studied. 
While varying $\beta$ (thus the temperature) we keep the physical
quark masses constant at $m_{ud} \approx 65$~MeV and $m_s \approx 135$~MeV
(the pion to rho mass ratio is $m_\pi/m_\rho\approx$0.66).
From now on we usually omit the different quark mass indices.

\begin{figure}
\begin{center}
\includegraphics*[width=8.5cm,bb=0 280 570 700]{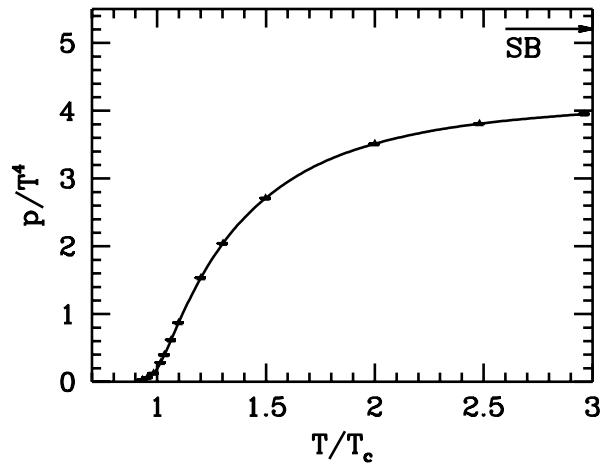}
\end{center}
\caption{\label{eos_p0}
The pressure normalised by $T^4$ as a function of $T/T_c$ at $\mu=0$ 
(to help the continuum interpretation the raw lattice result is multiplied with $c_p$).
The continuum SB limit is also shown.
}
\end{figure}
The determination of the equation of state at $\mu\neq$0
needs several observables ${\cal O}$
at non-vanishing $\mu$ values. This can be calculated by using  
the weights of eq. (\ref{reweight})
\begin{equation}
{\overline {\cal O}}(\beta,\mu,m)=\frac{\sum \{w(\beta,\mu,m,U)\} 
{\cal O}(\beta,\mu,m,U)}{\sum
\{w(\beta,\mu,m,U)\}}.
\end{equation}
We use the following notation for subtracting
the vacuum term: $\langle  {\cal O}(\beta,\mu,m) \rangle$= 
${\overline {{\cal {O}}}(\beta,\mu,m)}_{T\neq0}-
{\overline {{\cal O}}(\beta,\mu=0,m)}_{T=0}$.

\begin{figure}
\begin{center}
\includegraphics*[width=8.5cm,bb=0 280 570 700]{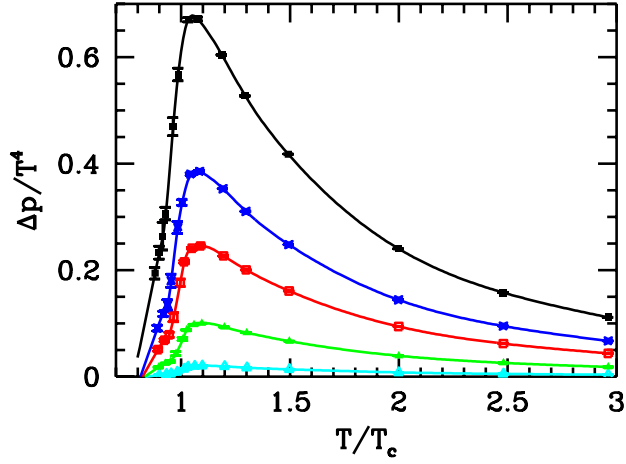}
\end{center}
\caption{\label{eosmu_p_sub}
$\Delta p=p(\mu\neq 0,T)-p(\mu=0,T)$ normalised by $T^4$
as a function of $T/T_c$ 
for $\mu_B$=100,210,330,410~MeV and
$\mu_B$=530~MeV (from top to bottom; 
to help the continuum interpretation the raw lattice result is multiplied with $c_\mu$).
}
\end{figure}

The pressure ($p$) can be obtained from the partition function
as $p$=$T\cdot\partial \log Z/ \partial V$ which can be written as
$p$=$(T/V) \cdot \log Z$ for large homogeneous systems.
On the lattice we can only determine the derivatives of $\log Z$ with respect
to the parameters of the action ($\beta, m, \mu$), so $p$ can be written as 
an integral\cite{Engels:1990vr}:
\begin{eqnarray}
\frac{p}{T^4}&=&\frac{1}{T^3 V} \int d(\beta, m,\mu ) 
\left(
\left\langle \frac{\partial(\log Z)}{\partial \beta}\right\rangle,
\left\langle \frac{\partial(\log Z)}{\partial m}\right\rangle,
\left\langle \frac{\partial(\log Z)}{\partial \mu }\right\rangle\right).
\end{eqnarray}
The integral is by definition independent of the integration path.
The chosen integration paths are shown on Fig \ref{weightlines}. 

The energy density can be written as 
$\epsilon =(T^2/V)\cdot \partial(\log Z)/\partial {T} 
+(\mu T/V)\cdot \partial(\log Z)/\partial\mu$.
By changing the lattice spacing $T$ and $V$ are simultaneously varied.
The special combination $\epsilon-3p$ contains only 
derivatives with respect to $a$ and $\mu$:
\begin{equation}
\frac{\epsilon-3p}{T^4}=-\left.\frac{a}{T^3V}\frac{\partial \log(Z)}{\partial a}\right|_\mu
+\left. \frac{\mu}{T^3 V}\frac{\partial(\log Z)}{\partial\mu}\right|_a.
\end{equation}

\begin{figure}
\begin{center}
\includegraphics*[width=8.5cm,bb=0 280 570 700]{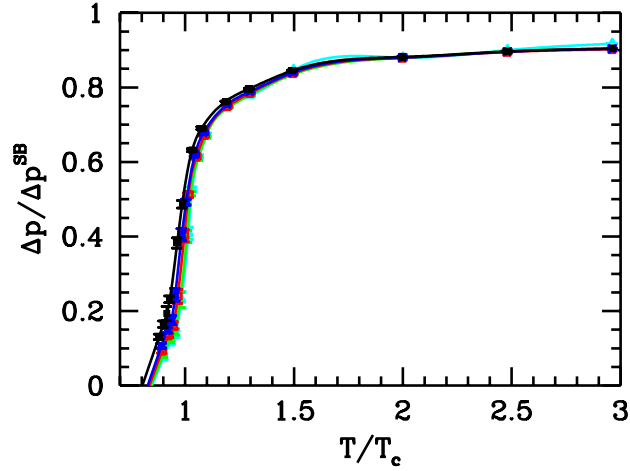}
\end{center}
\caption{\label{eosmu_p_SB}
$\Delta p$ of the interacting QCD
plasma normalised by $\Delta p$ of the free gas 
(SB) as a function of $T/T_c$ for $\mu_B$=100,210,330,410~MeV and
$\mu_B$=530~MeV. The result is essentially $\mu$-independent.
}
\end{figure}

The quark number density is $n=(T/V)\cdot \partial \log(Z)/\partial \mu$ which 
can be measured either directly or obtained from the pressure
(note, that the baryon density is $n_B$=$n$/3 and the baryonic chemical 
potential is $\mu_B$=3$\mu$).

We present lattice results on $p(\mu=0,T)$, 
$\Delta p(\mu,T)=p(\mu\neq 0,T)-p(\mu=0,T)$, $\epsilon(\mu,T)$-3$p(\mu,T)$ and 
$n_B(\mu,T)$. Our statistical errorbars are also shown. They are rather small, 
in many cases they are even smaller than the thickness of the lines. 
On the figures we multiply the lattice results with  
the dominant correction factors between $N_t$=4 and the continuum
in the $T\rightarrow\infty$ case:
$c_p=p(\mu=0,T\rightarrow\infty,\rm{continuum})
	/p(\mu=0,T\rightarrow\infty,N_t=4)$=0.518
and  $c_\mu=\Delta p(\mu,T\rightarrow\infty,\rm{continuum})
/\Delta p(\mu,T\rightarrow\infty,N_t=4)$=0.446. The well-known
continuum expressions in the $T\rightarrow \infty$ Stefan-Boltzmann (SB) 
case are $p(\mu=0,T\rightarrow\infty,\rm{continuum})=
(16+21n_f/2)\pi^2T^4/90$ and $\Delta p(\mu,T\rightarrow\infty,\rm{continuum})=
n_f\mu^2T^2/2+{\cal O}(\mu^4)$. This way the results presented on the
figures might be interpreted as continuum estimates and could be directly used
in phenomenological applications.
Direct $N_t$=4 lattice results can be 
obtained by dividing the presented values with these $c_p,c_\mu$ 
correction factors.

\begin{figure}
\begin{center}
\includegraphics*[width=8.5cm,bb=0 280 570 700]{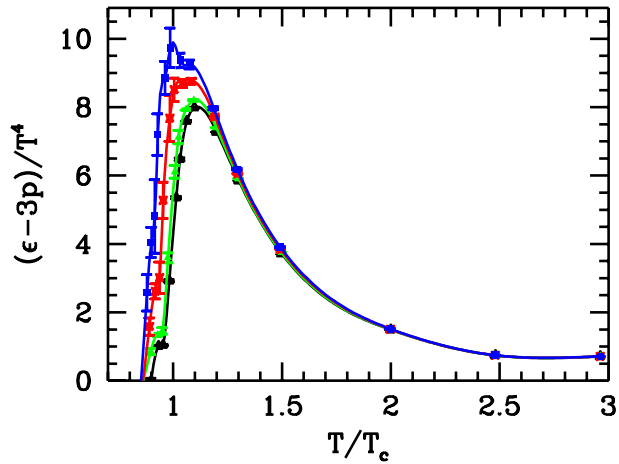}
\end{center}
\caption{\label{interaction}
$(\epsilon-3p)/T^4$ at $\mu_B$=0,210,410~MeV and
$\mu_B$=530~MeV as a function of $T/T_c$
(from bottom to top; 
to help the continuum interpretation the raw lattice result is multiplied with $c_p$)}.
\end{figure}

\begin{figure}
\begin{center}
\includegraphics*[width=8.5cm,bb=0 280 570 700]{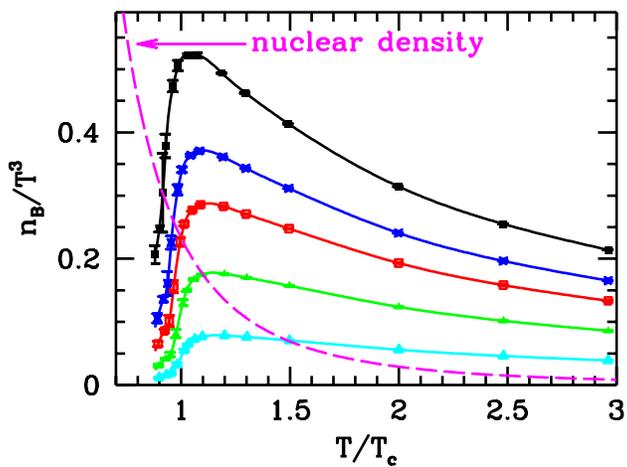}
\end{center}
\caption{\label{density}
The quark number density normalized $T^3$
as a function of $T/T_c$ for $\mu_B=$100,210,330,410~MeV and
$\mu_B$=530~MeV.
(to help the continuum interpretation the raw lattice result is multiplied with
$c_\mu$). As a reference value the line starting in the left upper corner 
indicates the nuclear density. 
}
\end{figure}

Fig. \ref{eos_p0} shows the pressure at $\mu$=0. 
On Fig. \ref{eosmu_p_sub} we present $\Delta p/T^4$ for five different 
$\mu$ values. 
Fig. \ref{eosmu_p_SB} gives $\Delta p(\mu,T/T_c)$ 
normalised by $\Delta p^{SB}=\Delta p(\mu,T\rightarrow\infty)$.
Notice the interesting scaling behaviour. 
$\Delta p / \Delta p^{SB}$ depends only on T and it
is practically independent of $\mu$ in the analysed region. 
Fig. \ref{interaction} shows $\epsilon$-3$p$ normalised by
$T^4$, which tends to zero for large $T$. 
Fig. \ref{density} gives the baryonic density
as a function of $T/T_c$ for different $\mu$-s. As it can be seen 
the densities exceed the nuclear density
by up to an order of magnitude.

As an important finding we mention that in the present analysis 
the applicability of our reweighting
method, the maximal $\mu$ value 
scales with the volume as 
$\mu_{\rm{max}}\cdot a \sim (N_t\cdot N_s^3)^{-0.25}$.
If this behaviour persists, one could --in principle--
approach the true continuum limit (note, that for 
$N_t\rightarrow\infty$ $a \sim 1/N_t \sim (N_t\cdot N_s^3)^{-0.25}$, thus
$\mu_{\rm{max}}$ is constant).

In this paper we studied the thermodynamical properties
of QCD at nonzero $\mu$. We used staggered QCD with 2+1 quarks on $N_t$=4
lattices. We determined for the first time the equation of state at nonzero
temperature and chemical potentials. Future analyses should
be performed at smaller lattice spacings and quark masses.
A detailed version of the present work will be published 
elsewhere \cite{Csikor:2002aa}.

\vspace{0.5cm}
\noindent
{\bf Acknowledgements:\\} 
We thank F.~Csikor, G.~Egri, I.~Montvay and 
A.A.~T\'oth for their help and suggestions.
This work was partially supported by Hungarian Scientific
grants, OTKA-T34980/\-T29803/\-M37071/\-OMFB1548/\-OMMU-708. 
For the simulations a modified version of the MILC
public code was used (see http://physics.indiana.edu/\~{ }sg/milc.html). 
The simulations were carried out on the 
E\"otv\"os Univ., Inst. Theor. Phys. 163 node parallel PC cluster.


\begin{thebibliography}{99}

\bibitem{Alford:1997zt}
M.~G.~Alford, K.~Rajagopal and F.~Wilczek,
Phys.\ Lett.\ B {\bf 422} (1998) 247
[arXiv:hep-ph/9711395].
 
\bibitem{Alford:1998mk}
M.~G.~Alford, K.~Rajagopal and F.~Wilczek,
Nucl.\ Phys.\ B {\bf 537} (1999) 443
[arXiv:hep-ph/9804403].
 
\bibitem{Rapp:1997zu}
R.~Rapp, T.~Schafer, E.~V.~Shuryak and M.~Velkovsky,
Phys.\ Rev.\ Lett.\  {\bf 81} (1998) 53
[arXiv:hep-ph/9711396].

\bibitem{Rajagopal:2000wf}
K.~Rajagopal and F.~Wilczek,
arXiv:hep-ph/0011333.

\bibitem{Fodor:2001au}
Z.~Fodor and S.~D.~Katz,
Phys.\ Lett.\ B {\bf 534} (2002) 87
[arXiv:hep-lat/0104001].

\bibitem{Csikor:2002aa}
F. Csikor et al., in preparation.

\bibitem{Hasenfratz:1983ba}
P.~Hasenfratz and F.~Karsch,
Phys.\ Lett.\ B {\bf 125} (1983) 308.
 
\bibitem{Kogut:1983ia}
J.~B.~Kogut, H.~Matsuoka, M.~Stone, H.~W.~Wyld, S.~H.~Shenker, J.~Shigemitsu
and D.~K.~Sinclair,
Nucl.\ Phys.\ B {\bf 225} (1983) 93.
                                                          
\bibitem{Montvay:1994aa}
I.~Montvay and G.~M\"unster, Quantum fields on a lattice.
Cambridge, UK, University Press (1994).        

\bibitem{Gottlieb:1996ae}
S.~Gottlieb {\it et al.},
Phys.\ Rev.\ D {\bf 55} (1997) 6852
[arXiv:hep-lat/9612020].

\bibitem{Karsch:2000ps}
F.~Karsch, E.~Laermann and A.~Peikert,
Phys.\ Lett.\ B {\bf 478} (2000) 447
[arXiv:hep-lat/0002003].
 
\bibitem{AliKhan:2001ek}
A.~Ali Khan {\it et al.}  [CP-PACS collaboration],
Phys.\ Rev.\ D {\bf 64} (2001) 074510
[arXiv:hep-lat/0103028].

\bibitem{Ejiri:2000bw}
S.~Ejiri,
Nucl.\ Phys.\ Proc.\ Suppl.\  {\bf 94} (2001) 19
[arXiv:hep-lat/0011006].

\bibitem{Ferrenberg:yz}
A.~M.~Ferrenberg and R.~H.~Swendsen,
Phys.\ Rev.\ Lett.\  {\bf 61} (1988) 2635.

\bibitem{Barbour:1997ej}
I.~M.~Barbour, S.~E.~Morrison, E.~G.~Klepfish, J.~B.~Kogut and M.~P.~Lombardo,
Nucl.\ Phys.\ Proc.\ Suppl.\  {\bf 60A} (1998) 220
[arXiv:hep-lat/9705042].

\bibitem{Fodor:2001pe}
Z.~Fodor and S.~D.~Katz,
JHEP {\bf 0203} (2002) 014
[arXiv:hep-lat/0106002].

\bibitem{Berges:1998rc}
J.~Berges and K.~Rajagopal,
Nucl.\ Phys.\ B {\bf 538} (1999) 215
[arXiv:hep-ph/9804233].

\bibitem{Halasz:1998qr}
M.~A.~Halasz, A.~D.~Jackson, R.~E.~Shrock, M.~A.~Stephanov and
J.~J.~Verbaarschot,
Phys.\ Rev.\ D {\bf 58} (1998) 096007
[arXiv:hep-ph/9804290].

\bibitem{Allton:2002zi}
C.~R.~Allton {\it et al.},
arXiv:hep-lat/0204010.

\bibitem{deForcrand:2002ci}
P.~de Forcrand and O.~Philipsen,
arXiv:hep-lat/0205016.

\bibitem{Engels:1990vr}
J.~Engels, J.~Fingberg, F.~Karsch, D.~Miller and M.~Weber,
Phys.\ Lett.\ B {\bf 252} (1990) 625.

\end{thebibliography}
\end{document}